**Structure, Dynamics and Hydrogen Transport in Amorphous Polymers: An Analysis of the Interplay Between Free Volume Element Distribution and Local Segmental Dynamics from Molecular Dynamics Simulations**

Mohammed Al Otmi[a], Frank Willmore[b], Janani Sampath[a,*]

[a]Department of Chemical Engineering, University of Florida, Gainesville, Florida 32611

[b]Department of Research Computing, Boise State University, Boise ID 83725

*jsampath@ufl.edu

**Abstract**

Polymers are attractive membrane materials owing to their mechanical robustness and relatively inexpensive fabrication. An important indicator of membrane performance are free volume elements (FVEs) – microporous void spaces created by the inefficient packing of bulky groups along the polymer chain. FVEs tend to degrade over time as polymer chains reorganize irreversibly. While it is widely accepted that polymer flexibility has an impact on membrane transport properties, the molecular nature of this impact is still not well-understood. By establishing a correlation between local chain dynamics and the distribution of free volume elements (FVEs), penetrant transport can be regulated more efficiently in amorphous polymer membranes. In this work, we implement all-atom molecular dynamics (MD) simulations to explore the relationship between chain dynamics and free volume in three polymers with different levels of backbone flexibility— polymethylpentene (PMP), polystyrene (PS), and HAB-6FDA thermally rearranged polymer (TRP). We construct these polymers at different temperatures and examine how temperature impacts the FVE distribution and segmental mobility. Our analysis shows that chain segments near FVEs have higher mobility compared to the atoms in the bulk; the extent of this difference increases with chain flexibility. Increasing chain flexibility by increasing the temperature results in a broader FVE distribution. Rigid polymers such as TRP show the most robust FVE distribution and are not significantly affected by temperature change. To capture penetrant diffusion through the polymer matrix, hydrogen is inserted, and the diffusion is measured at different temperatures; hydrogen mobility is influenced by the FVE structure and overall mobility of polymer chains. At low temperatures, hydrogen mobility is influenced by void distribution, while at high temperatures, polymer dynamics dictates hydrogen transport.

**Introduction**

Traditional hydrocarbon purification is carried out through cryogenic distillation, which requires a large amount of energy.[1] Integrating membrane-based separation with distillation is a promising alternative to reduce the energy consumption in the petrochemical industry. The performance of a membrane is characterized by the amount of penetrant flux or permeability, and the efficiency of separation or selectivity. For gas separation membranes, these factors are inversely correlated, as illustrated by the permeability-selectivity plot developed by Robeson.[2] Amongst the different membrane-based technologies that are available today, polymers are attractive materials due to their mechanical robustness, tunable properties, and inexpensive means of fabrication.[3–5] Despite their many benefits, polymer membranes are limited in their application due to the inherent flexibility that arises from the molecular conformation of the chains in the matrix. This flexibility leads to structural rearrangement that causes membranes to lose their efficiency over time, due to phenomena such as plasticization and physical aging.[5–8] A grand challenge is to develop polymers that are tailored to separate gases while maintaining a high level of selectivity with long lifetimes.[1]

Amorphous polymers below their glass transition temperature ($T_g$) that have rigid chains and form microporous networks are superior membrane materials compared to their rubbery counterparts.[4,9] An important feature of glassy polymers are free volume elements (FVEs) – the distribution of highly microporous void spaces created by the inefficient packing of bulky groups along the polymer backbone.[10,11] While the voids themselves drive the thermodynamic solubility of penetrant, the shape and connectivity of these voids are the primary driver of transport in glassy polymers. Due to this network of voids, the primary driving force for separation in glassy polymers is penetrant diffusivity, with solubility playing a smaller role in the transport of penetrants through the membrane.[12,13] The upper bound of the Robeson plot is populated with highly glassy polymers mainly due to their superior diffusion selectivity.[14–16] The free volume theory of penetrant transport posits that the diffusion of a penetrant through the polymer matrix occurs through a sequence of hops between voids that have sufficient space to accommodate it. Free volume theory attributes the rate and mechanism of gas diffusion in polymers to the size and distribution of FVE spaces in the polymer matrix.[13,17,18] The correlation between FVE and diffusivity of gas molecules in glassy polymers makes FVEs an important parameter to consider when designing membrane materials.[18–20]

Having a robust FVE distribution is necessary to maintain polymer performance at long lifetimes. While the relative stability of FVEs in glassy polymers makes them superior separation media, they still experience structural relaxation and chain reorganization which compromises their performance.[21,22] This stems from the fact that although polymers have restricted dynamics below their glass transition temperature ($T_g$), they are not completely immobile.[23,24] Under certain operating conditions (high pressures and temperatures, and high penetrant concentrations) and under ambient conditions as well, their chains become susceptible to rearrangements which alter the polymer morphology and FVE size and distribution. Local segmental motions, including translational, vibrational, and rotational motions of the chains, influence the FVE structure. FVEs can form, collapse, or coalesce due to thermally activated chain movement.[25] Moreover, it is important to consider the coupling between chain segmental motion and FVEs in facilitating penetrant transport when designing novel membrane materials.

Physical aging and plasticization are two phenomena preventing the utilization of hydrocarbon separation membranes on a commercial scale, both of which are related to chain flexibility. Physical aging involves structural relaxation in glassy polymers under ambient conditions, driven by the difference in the current specific volume of the membrane and its equilibrium specific volume.[6] This leads to the dissipation of nonequilibrium free volume and the densification of the membrane, causing the pores in the polymer structure to collapse, reducing the overall permeability.[6,26] On the other hand, plasticization arises from irreversible chain reorganization of the polymer matrix and causes membrane selectivity to decline over time.[27,28] It occurs when low-molecular-weight molecules adsorb onto the interstitial spaces of the polymer matrix at high pressures. The polymer chain becomes more flexible, and the interchain spacing increases, reducing the membrane's sieving capability.[5,6] Both these interrelated processes are directly correlated to the flexibility of the polymer chains, and their ability to undergo microscopic chain reorganization. Elucidating the molecular underpinnings of chain rearrangement and the relationship between chain dynamics and FVE structure will enable the rational design of durable, high-performance membranes.

Common FVE characterization techniques such as positron annihilation lifetime spectroscopy (PALS), NMR spectroscopy, and photochromic probing have been used to obtain the average FVE distribution in a polymer membrane at a given point in time.[29–31] Although these methods are powerful, they cannot directly capture the transient nature of FVEs that arises due to segmental motions of the chains. Recently, restricted orientation anisotropy method (ROAM) has been shown to provide a measure of the dynamics of FVE by correlating the orientational confinement experienced by the probe molecule to the local structural environment.[10,11] While we can gain insights about FVE distribution using these techniques, the underlying molecular structure-property relationships (change in FVE as a function of temperature or chain flexibility for instance) are hard to probe directly using experimental methods alone.[32]

Over the years, molecular simulations have successfully captured penetrant adsorption and transport through different amorphous polymers. Specifically, the Colina group has studied the adsorption of penetrants through polymers of intrinsic microporosity (PIMs) using a combination of grand canonical Monte Carlo (GCMC) and MD simulations, revealing complex links between matrix swelling, free volume element rearrangement, and sorption-induced adsorption of penetrants at different pressures.[32–35] Kumar et al. have used coarse-grained models to study the relationship between penetrant size and polymer matrix motion, as well as the mechanism of gas diffusion in polymer grafted nanoparticle membranes.[36–38] Additionally, several other modeling studies have elucidated the molecular nature of gas and liquid transport in amorphous polymers. [33,39–43] Overall, the relaxation of amorphous polymers like polystyrene and poly(1-trimethylsilyl-1-propyne) has been studied using different molecular modeling techniques, including coarse-grained and atomistic methods. [44–46] However, limited work has been done correlating the free volume elements and segmental dynamics in state-of-the-art rigid polymers like thermally rearranged polymers, that have shown promise as hydrocarbon separation membranes.

In this work, we utilize all-atom molecular dynamics (MD) simulations to create three distinct amorphous polymers, namely polymethylpentene (PMP), polystyrene (PS), and HAB-6FDA thermally rearranged polymer (TRP). These polymers exhibit diverse chemical structures, with TRPs belonging to a class of novel polymers with tunable pore distributions. Amorphous polymers below their $T_g$ are known to have a long relaxation time which makes it challenging to characterize their dynamics in classical MD timescales. To investigate the trends in flexibility, it is necessary to sample the systems at elevated temperatures, where the mobility of the polymer can be distinctly captured, hence we simulate these systems at three temperatures ranging from 100 K to 500 K. We probe the structure of these systems using two different free volume element measurements. Next, we study segmental dynamics by calculating the bond autocorrelation function. To investigate the interplay between chain flexibility and void distribution on diffusion properties, we randomly introduce hydrogen molecules into the three polymer systems and analyze the diffusion behavior of hydrogen with respect to polymer chemistry and temperature. Quantifying the dynamics of different segments of the polymers with respect to FVEs can be used to elucidate the potential for a polymer to undergo rearrangement and deviate from its initial FVE distribution when used as a membrane material.

**Methods**

To model the polymer systems, we utilize the OPLS-AA forcefield to represent the interactions in the three polymer chemistries. OPLS-AA parameters are taken from LigParGen web server which uses 1.14 CM1A and 1.14 CM1A-LBCC charge models.[47] After parameterizing and optimizing the monomer structure, we pack multiple monomers in a simulation box at a low density. We use the open-source package Polymatic, which has been used to generate a wide variety of polymers in the past, to build the polymer chain.[48] Polymatic uses the simulated polymerization algorithm to construct the polymer chains; details can be found in the Supporting Information. Briefly, bonds are formed between monomer units to form a single polymer chain. Before the addition of each bond, energy minimization is performed to relax the newly formed bond. Short MD simulations in the NVT and NPT ensembles are performed after every fifth bond in order to perturb the system and allow polymerized fragments to connect. After the formation of a single chain, twenty chains are packed into a large simulation box following which a 21-step equilibration protocol is performed to bring the systems to the appropriate density at the desired temperature and 1 atm.[49] This protocol uses a number of NPT and NVT ensembles to heat and compress the system to a high temperature and pressure, then allows the system to cool and decompress to the final condition gradually to avoid any artifacts associated with the stress tensor.[50] Simulated samples of PMP and PS consist of 200 monomers per chain while TRP consists of 100 monomers per chain. PS is atactic, with randomly distributed isotactic and syndiotactic segments. After the 21-step equilibration, we run the systems in the isothermal-isobaric ensemble at the specified temperature (100 K, 300 K, 500 K) and 1 atm for 3 ns to ensure that the systems fluctuate within their set densities.[51,52] To validate the forcefield, we examine the density and glass transition temperature ($T_g$) of the three structures. We compute $T_g$ of the three systems using a gradual quenching protocol.[53,54] The details and results of the protocol are presented in the Supporting Information.[55] The calculated $T_g$ for PMP, PS, and TRP were 331K, 407K and, 579K, respectively, which agree with experimental $T_g$ reported in the literature, of 300K, 373K, and 578K.[53,29,30]

The size of our simulation box ranges ~700 nm$^3$ to ~1400 nm$^3$, and they each consist of 64000 – 106000 atoms. While performing replicas in MD simulations is useful for increasing the overall sampling and accounting for uncertainty, our systems are large enough to obtain a representative sample of the system's behavior, and sufficient sampling is generated from a single large system. However, to ensure that our systems are sufficiently large, we run three independent replica simulations for the three chemistries at 300 K and examine the trends in segmental dynamics. This comparison is included in the Supplementary Information and shows that the trends observed are consistent across different replicas.

Production runs are performed at different temperatures (100K, 300K, 500K) and 1 atm for 20 ns, over which data is collected. Nose-Hoover thermostat and barostat are employed, with a timestep of 1fs, and temperature and pressure damping parameters of 100 ps and 1000 ps, respectively. To study penetrant diffusion, hydrogen gas molecules are randomly inserted into the three equilibrated polymer structures at the different temperatures such that it corresponds to a gas concentration of 1.18 mol/L, equivalent to hydrogen molecular density at 29.6 atm and 300 K. After this, a short equilibration of 1 ns followed by a 20 ns production run was performed in NVT, similar to the method outlined in prior work.[42,43] This insertion protocol bypasses the time needed to achieve a penetrant-saturated configuration. All simulations are carried out using the LAMMPS package.[56] Long range interactions are calculated using particle-particle particle mesh algorithm (PPPM) with a 1.5 nm cutoff.

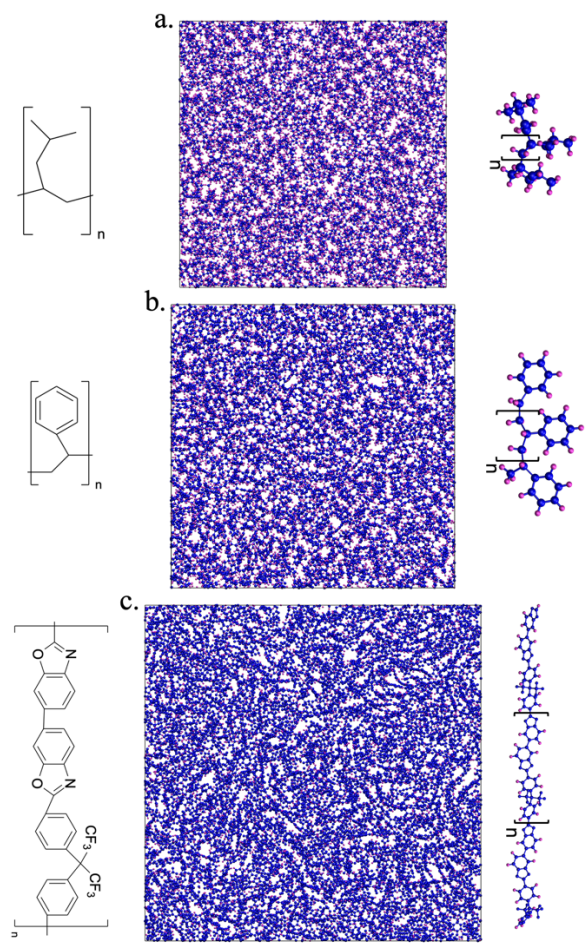

**Figure 1.** Chemical structures and snapshots of the three polymer systems – (a) Polymethylpentene (PMP), (b) Polystyrene (PS), (c) HAB-6FDA TR (TRP). Chains are constructed using Polymatic. Each system has 20 chains; PMP and PS contain 200 monomers per chain, TRP contains 100 monomers per chain. Heavy atoms are shown in blue, and hydrogens are in pink. Snapshots depict 2.5 nm slices in the z direction for each system at 300 K.

## Results

### Free Volume Element Distribution

To probe the structure of the three polymers, we assess their microporosity by calculating the free volume of the polymer matrix. Free volume in polymers can be defined in different ways; a summary of these definitions and their differences can be found in the detailed perspective by White and Lipson.[57] Specifically, we calculate the total free volume, which is the sum of the occupied volume, and excess free volume. Computationally, free volume can be obtained by energetic or geometric methods.[58–61] Each method of analysis sheds light on a certain aspect of void distribution; taken together, they provide a complete description of the microstructure of the polymer. The purpose of this study is to compare the change in free volume as a function of chain architecture and temperature, not to correlate free volume with $T_g$. In the two methods of measurement that we employ, we remain consistent in the parameters that we choose across the different chemistries and temperatures.

To calculate the overall void volume, we utilize the alpha-shape method within the molecular visualization package Ovito.[62] Alpha-shape is a geometric method of FVE estimation in which the particle coordinates are used to tessellate the simulation box into tetrahedral elements after which a probe sphere with pre-defined radius is employed to classify these elements into either empty or filled regions. The probe size is directly correlated to the void volume in the polymer – smaller probes yield large free volumes, whereas a large probe results in smaller total free volume. The alpha-shape method also allows for the classification of atoms as a function of their distance from FVEs.[63] Particularly, regions within a surface manifold bound by the intersection between empty regions (voids) and filled regions (bulk polymer) can be identified. In this way, we classify polymer atoms as belonging to the surface (near voids) or bulk (away from the voids) depending on their atomic positions at t = 0. This allows us to capture the structural heterogeneities of the polymer and draw comparison between the dynamics in surface and bulk regions of different polymers, as a function of temperature.

A common criterion of choosing probe size is to use a value higher than the nearest neighbor atom separation distance. The probe radius employed (0.25 nm) was chosen to be higher than the atomic distance between two carbon atoms along a polystyrene backbone. As seen in Figure 2, the void volume (ratio of free to total volume) in the three systems increases as PMP > PS > TRP. This is expected, as TRP monomers have multiple cyclic groups along the backbone (Figure 1c), hindering chain packing, thereby increasing free volume. This is also in line with the simulated $T_g$ measurements – polymers with high $T_g$ are stiffer and have high free volumes; TRP has the highest $T_g$ while PMP has the lowest. We also find that for each polymer, the void volume increases with increasing temperature, in line with free volume and glass transition theories that predict an increase in free volume as a function of temperature.[57,64] Moreover, PMP shows the highest increase in free volume between the lowest and highest temperatures considered – this is because PMP has the highest flexibility and lowest $T_g$. TRP on the other hand shows a modest change in free volume with temperature. We note that the free volume (void ratio) of TRP is higher than free fractional volume (FFV) reported in prior simulations,[65] this is due to the probe size used – we have chosen a probe radius that generates statistically significant number of surface atoms that allows us to calculate local dynamics across all three systems at the different temperatures considered. A larger probe (0.3 nm) significantly reduces the void ratio from 43% to 25%. We also report the densities of the three systems at different temperatures in the Supporting Information. De-densification of the polymer could contribute to some changes in the free volume as well.

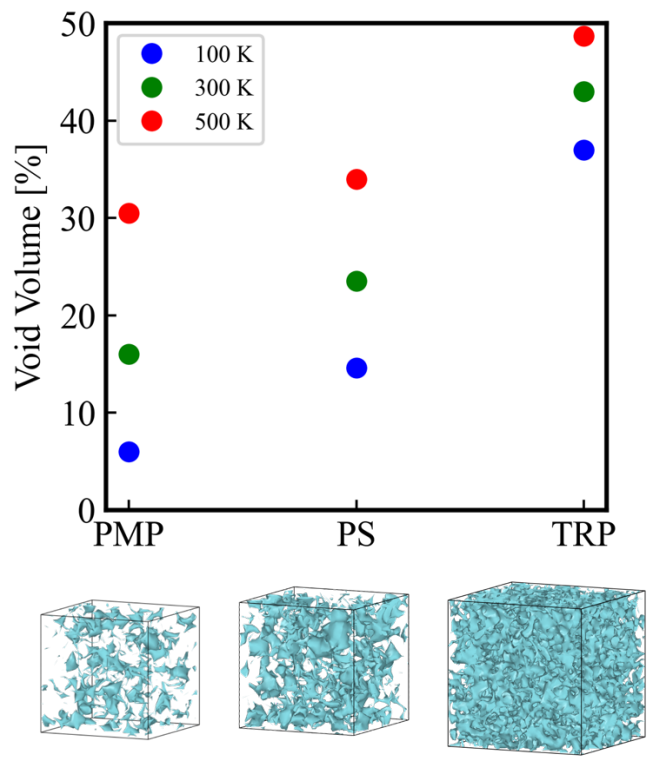

**Figure 2.** Void volume across the three polymer systems, at three different temperatures, as labelled. Voids are obtained by constructing a surface mesh on Ovito using the alpha-shape method, with a spherical probe of 0.25 nm radius. Snapshots of voids at 300 K are shown directly below each system along the x-axis.

To understand the distribution of FVEs across the different systems, we employ the Void Analysis Codes and Unix Utilities for Molecular Modeling and Simulations (VACUUMMS).[66] This software package implements the Cavity Energetic Sizing Algorithm, a Monte Carlo based energetic method of determining voids in a polymer matrix.[67,68] In CESA, a cavity is defined as a spherical volume with an energy center which is the local minimum in a repulsive particle energy field. This gives rise to a size distribution of voids. CESA has been shown to correlate well with permeability properties and PALS measurements.[69] Figure 3 shows qualitative snapshots (rendered from scene definition language (SDL) generated by VACUUMMS using POV-Ray[70] tracing software) of the FVEs in PMP, PS, TRP, across three different temperatures. We find that temperature plays a critical role on the pore size distribution, and not only the overall free volume distribution, as seen in Figure 3. We see that there is an increase in the size of the individual FVEs as the temperature increases; this is most pronounced in PMP, followed by PS, and to a lesser extent in TRP.

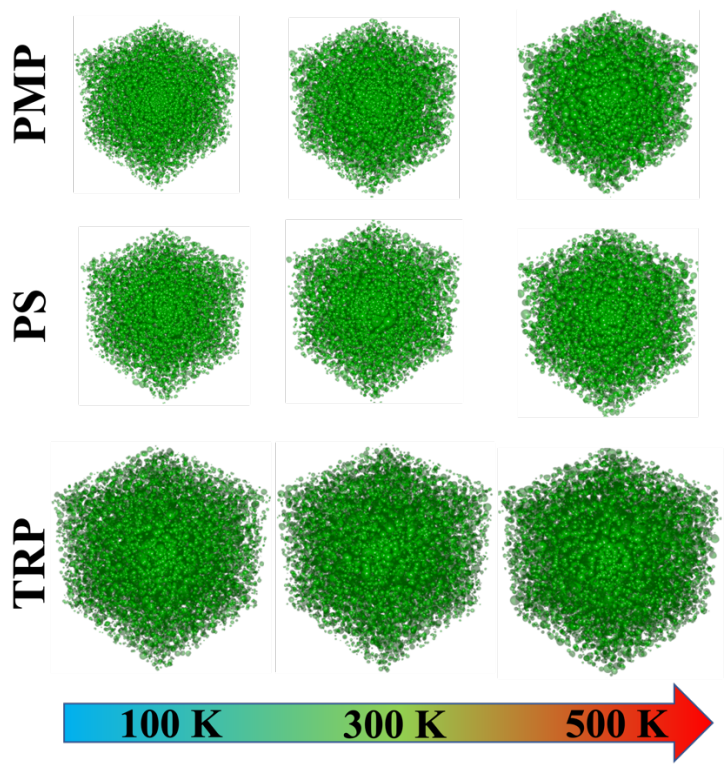

**Figure 3. Cavities as a function of temperature, for the three different polymer systems, as labeled. Images rendered using POV-Ray, from data generated via VACUUMMS.**

To quantify the voids, we plot the cavity size distribution, as seen in Figure 4. In agreement with the snapshots, we see that there is a shift towards larger cavity sizes with an increase in temperature, which is most pronounced in PMP. It is also seen that at the lowest temperature, the voids in PMP are the smallest, followed by PS and TRP, however, at the highest temperature, PMP has a higher distribution of large voids. Given the flexibility of polymers, thermal fluctuations cause pores to coalesce at higher temperatures; this is most distinct in systems that are flexible, compared to systems with stiffer chains. This is why PMP exhibits the greatest sensitivity to temperature variations, as evidenced by the significant shift in the distribution of FVE observed between 100 K and 500 K. PS experiences a moderate alteration in its FVE distribution, marked by a shift in the distribution towards pores with larger diameters, and pores in TRP do not change very much in the temperature range considered (although there is a slight shift toward larger pores at 500 K). This also correlates with the $T_g$ of these systems – the $T_g$ of TRP is greater than 500K, thus, chain mobility is restricted at this temperature, unlike PMP and PS. Overall, we find that as temperature increases, the number of small-diameter voids decreases while that of larger-diameter voids increases. This observation suggests that, in addition to the established phenomenon of thermal expansion of free volume,[57,71] the small individual pores within the polymer matrix coalesce to produce larger pores.

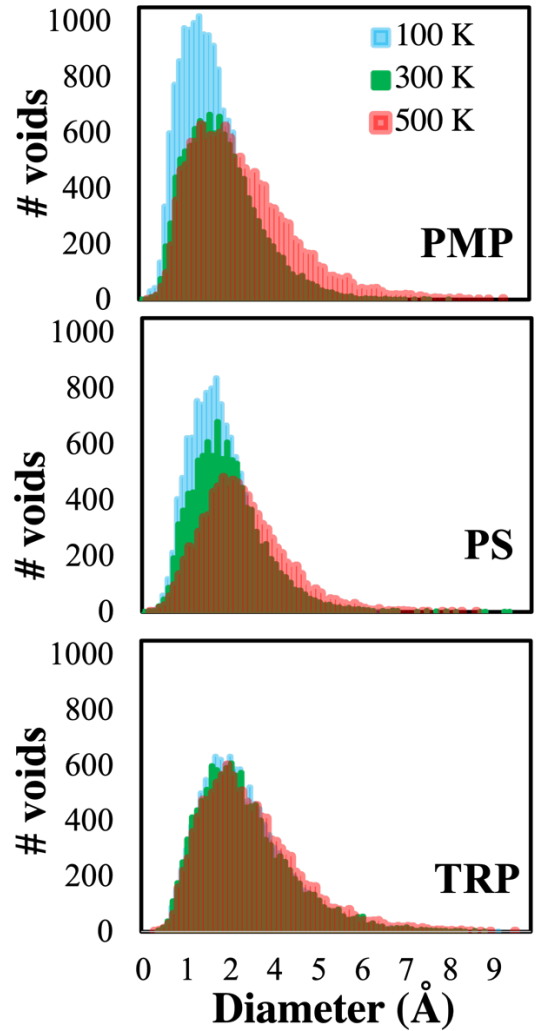

**Figure 4:** Cavity size distribution for (a) PMP (b) PS and (c) TRP as a function of temperature, as labelled

**Polymer Dynamics**

Next, we evaluate the translational dynamics of the three polymers at the different temperatures by calculating the mean squared displacement (MSD). MSD measures the change in the position of the particles from their initial positions, defined as:

$$MSD\ (t) = <\Delta r^2(t)> = <(r(t_0 + t) - r(t_0))^2>$$

where r ($t_0$ + t) is the position of a particle at time t. We do not examine the center of mass motion of each chain, but instead compute the dynamics of all atoms (including hydrogen atoms along the polymer backbone and sidechain) across the three systems to improve statistics.

Universally, polymers exhibit three distinct regions in the mean-squared displacement.[23,24,72] At short times (< 1 ps), all three systems follow ballistic dynamics (slope of 2.0). After about 1 ps, the dynamics of the three systems move into the sub-diffusive regime (slope < 1.0). A distinct plateau regime develops after 1 ps which is more prominent in PS and TRP, presumably due to the time-scale separation between vibrational and relaxational degrees of freedom, restricting the dynamics of the atoms.[23,24] This plateau regime persists beyond 1000 ps at 100 K and 300 K, and at 500 K, it only persists till 10 ps. At long times, polymer melts and rubbers attain diffusive behavior; however, none of the three systems at temperatures < 500 K reach diffusive regime due to the restricted dynamics and constrained nature of the chains. At 500 K, PMP atoms are diffusive at long times. The trends shown for MSD have been observed in other amorphous systems.[23,24] At any given temperature, PMP exhibits faster dynamics compared to PS and TRP, which is expected given that PMP has the lowest $T_g$ of the three polymers considered, and a more flexible backbone. Additionally, increasing the system temperature accelerates the overall dynamics of each system, which can be seen in the larger MSD values at elevated temperatures.

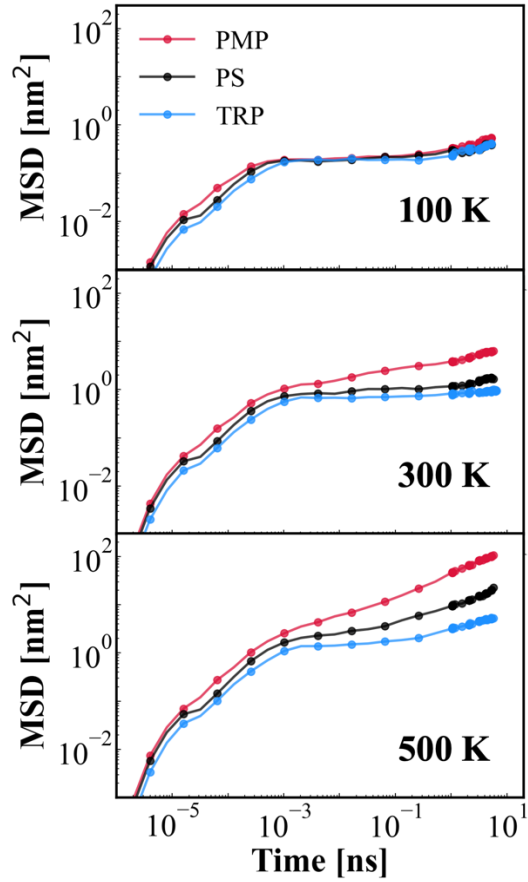

**Figure 5: Mean squared displacement of all polymer atoms in the system for PMP, PS, TRP at 100 K, 300 K, 500 K, as labelled.**

To study the orientational dynamics and to probe relaxations on a more local scale, we calculate the bond vector autocorrelation function (bond ACF). It is defined by

$$\text{Bond ACF} = \left\langle \frac{\overrightarrow{B(t)} \cdot \overrightarrow{B(0)}}{\overrightarrow{B(0)} \cdot \overrightarrow{B(0)}} \right\rangle$$

where B(t) is bond vector at time t. Averaging is done over all bonds in the system. Calculating the bond ACF enables us to capture small-scale orientational fluctuations in these amorphous systems. As seen in Figure 6, the rate of decorrelation of the bond ACF reveals a consistent pattern, with PMP exhibiting the fastest relaxation at any given temperature, followed by PS, and TRP displaying the slower relaxation. The influence of temperature on the bond ACF is similar to the trends observed in the MSD; as temperature increases, the relaxation of the bond segments is more rapid.

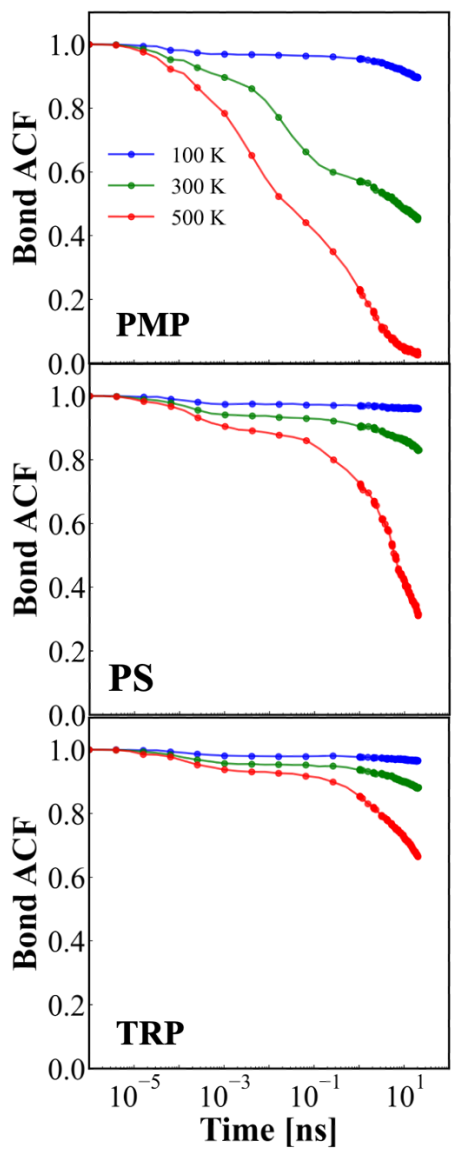

**Figure 6**: Bond autocorrelation function averaged over all bonds in the three systems at three different temperatures, as labelled.

Amorphous polymers are structurally heterogeneous, with voids interspersed between filled polymer regions. To probe polymer dynamics as a function of proximity to voids, we use the alpha-shape method to identify surface and bulk atoms (described above). Examining the dynamics of polymer segments that lie at the interface between void and filled regions offers a more robust correlation between the distribution of the FVE and the segmental dynamics. We compute the bond vector autocorrelation function of the surface and bulk atoms separately to evaluate the difference in dynamics between different regions in each of the three systems as shown in Figure 7.

We observe increased dynamics in the atoms that are at the surface, compared to atoms in the bulk, as evidenced by the faster relaxation of bond ACF of surface bonds compared to bulk bonds. The increased dynamics of the surface segments can be attributed to the greater number of degrees of freedom for these atoms, as a result of their proximity to a free interface. Our results also reveal that the contrast in dynamics between surface and bulk atoms is more distinct in PMP compared to PS and TRP. This can be attributed to the inherent rigidity of PS and TRP chains, regardless of their position in the bulk or near the surface. Conversely, the flexible chains in PMP exhibit a marked difference in dynamics between the bulk and surface regions. Atoms in the bulk region exhibit significantly slower dynamics due to the increased number of neighboring atoms that restrict their movement. The observed correlation between the dynamics of surface segments and the change in free volume element (FVE) distribution is ascribed to the fact that the surface segments constitute the walls of FVE, and thus their mobility directly influences the alteration in FVE. The dynamics of the surface segments offer an indication of the susceptibility of FVE distribution to change. The relatively slower dynamics in TRP compared to PMP suggests a more robust FVE distribution in TRP. We find that for the timescales considered here, there is no difference in dynamics between surface and bulk groups at 100 K for PS and TRP, and the difference in dynamics for PMP segments is noticeable only at long times at 100 K. Interestingly, for PMP the bulk vs. surface groups at 300 K show a greater difference in segmental dynamics than that at 500 K. This could be because the total segmental mobility is very high for PMP at 500 K, and the proximity to voids does not speed up the relaxation as much as it does at 300 K, where the overall mobility is restricted, thus the presence of voids causes a sharp increase in dynamics. A similar trend is observed for the PS and TRP systems at 500 K. While the PS segments show faster relaxation compared to TRP, we see that the difference in dynamics between surface and bulk bonds at 500 K is more pronounced in TRP than it is in PS. This is attributed to the fact that the overall high chain mobility reduces the impact of the void interface in PS, however, in TRP, the presence of voids greatly enhances the dynamics of chain segments that are close to it.

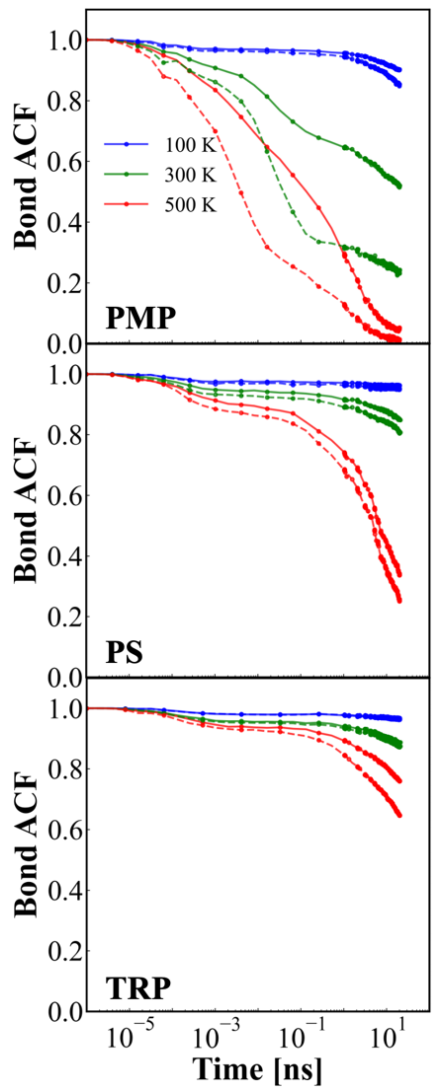

**Figure 7. Comparison between segmental orientational dynamics of surface particles and bulk particles. Solid lines show the Bond ACF for atoms in the bulk while dashed lines show bond ACF for surface atoms (atoms near the FVE) atoms near the FVEs.**

**Penetrant Transport**

To investigate the impact of polymer structure on penetrant diffusion in the three systems, hydrogen molecules were introduced, and their trajectories were analyzed to gain insight into the nature of molecular transport under different temperature conditions. First, we investigate hydrogen transport qualitatively over 300 ps, as shown in Figure 8. For each system, we find that hydrogen diffusion increases as temperature increases, as expected. Prior work by Brownell et. al has found Arrhenius dependence between hydrogen diffusion and temperature in amorphous polymers.[42] At a temperature of 100 K, hydrogen molecules are almost completely confined in PMP and PS, with only modest mobility observed in TRP. This is due to the high microporosity in TRP compared to the other systems at 100 K, which facilitates the transport of hydrogen. At 300 K, the hydrogen trajectories in PMP and TRP are similar (evidenced by the area of the box covered by hydrogens over this time). Both systems show higher hydrogen mobility than that of PS. At 500K, significantly higher hydrogen diffusion is observed in PMP, in comparison to PS and TRP. Overall, we find that the trends in hydrogen diffusion across the three systems is different at different temperatures.

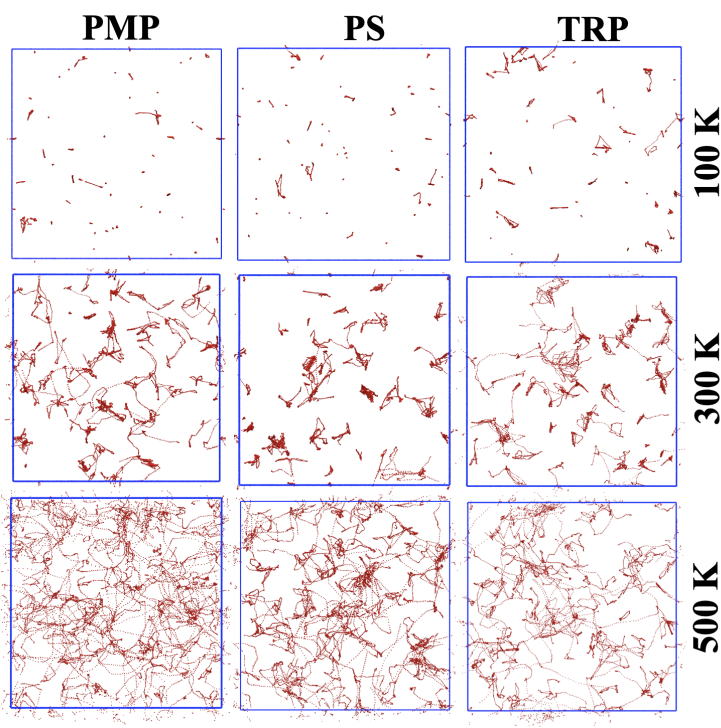

**Figure 8: Trajectories of 50 hydrogen molecules (chosen randomly) in PMP, PS and TRP at three different temperatures, as labelled. The evolution is shown from 0 ps to 300 ps. Each bead has a simulation dt of 10 ps.**

To investigate this further, we calculate the mean-squared displacement of all hydrogen atoms in the three systems at different temperatures (Figure 9). Across each system, there is an increase in hydrogen diffusion as the temperature increases, in line with prior work.[42] However, comparing hydrogen MSD of the three systems at each temperature reveals interesting trends. At low temperatures (100 K), TRP exhibits substantially higher hydrogen diffusion than PS and PMP. This can be ascribed to the elevated microporosity of TRP, enabling hydrogen molecules to hop between the voids. However, as the temperature increases to 300 K, we see a reversal in this trend. At short times, $H_2$ diffusion is highest in TRP, followed by PS and then PMP, however, at around 100 ps, the MSD of $H_2$ in PMP surpasses that of TRP. At 500 K, $H_2$ diffusion is highest in PMP, followed by PS and TRP.

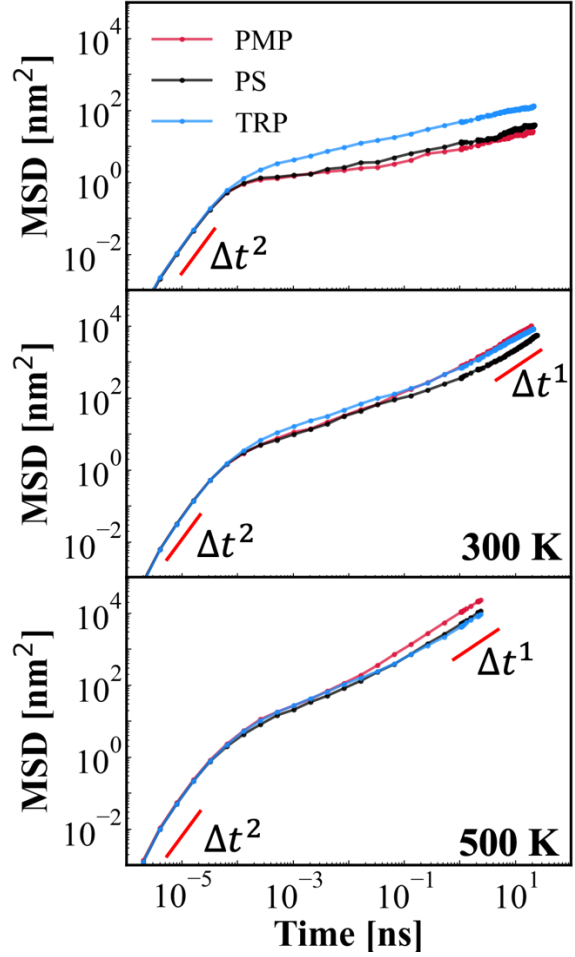

**Figure 9: Mean squared displacement of hydrogen molecules in PMP, PS and TRP at three different temperatures, as labeled.**

To provide additional insight about hydrogen dynamics, we compute the ensemble-averaged displacement of hydrogen molecules using the self-part of Van Hove correlation function, given by:

$$G_s(r,t) = \frac{1}{N} \langle \sum_{i=1}^{N} \delta(r - |r_i(t) - r_i(t_0)|) \rangle$$

where G(r,t) is the probability distribution, r is the position, t is the time, $\delta$ is the Dirac delta function, and N is the number of particles. Assuming that the particle is initially located at the origin at time t=0, the function G(r,t) denotes the probability of detecting the particle at position r within a time interval of dt. Figure 10 illustrates the dynamics of hydrogen in the three different systems and at different temperatures. Three distinct time intervals or dt = 1, 10 and 100 ps have been selected to demonstrate changes in hydrogen mobility at different times. We find that at a given temperature, the peak height decreases with increasing dt, signifying that hydrogen molecules have been displaced from their initial positions. We also see that for a given dt value, the peak height is lower at higher temperatures. At 100 K and 300 K, the displacement value of the first peak is constant for each system at a specific temperature, and a shoulder appears at long times (> 10 ps). This implies that at these temperatures, there are two different types of hydrogen molecules that contribute to the distribution– localized and mobile. Specifically at 300 K, the first and second peaks in the distribution at dt = 100 ps signify local and mobile hydrogen molecules, respectively, across all three polymers. At 500 K, there is a single distribution at dt = 100 ps, with almost all hydrogen molecules exhibiting similar dynamics, and the peaks have different displacement values, unlike the distributions at lower temperatures.

At 100 K, most hydrogens remain localized, resulting in small changes in peak height for PMP and PS. Additionally, the displacement remains nearly constant across different time intervals due to the limited mobility of hydrogen molecules. The peak height in TRP shows the most variation with dt, with a larger shift toward longer displacements, signifying that hydrogen dynamics is highest in TRP at 100 K. This is in line with the MSD results, as well as the fact that TRP has the largest void distribution at 100 K (Figure 3). At 300 K and dt = 100 ps, PS has the highest population of localized hydrogens and PMP has slightly more mobile hydrogens than TRP. This is consistent with the MSD results, with hydrogens in PS having lower MSD than that of PMP and TRP. At 500 K, the shoulder is not observed for PMP and PS, however, there is a faint shoulder in the TRP distribution at dt = 100 ps, indicating that hydrogen mobility is restricted in TRP at high temperatures, compared to PMP and PS.

The increase in hydrogen mobility in PMP at higher temperatures can be attributed to two factors. The first is that the PMP chains are inherently more mobile compared to TRP and PS, this mobility contributes to the penetrant motion. However, this is predominant only at high temperatures. At lower temperatures, when chain mobility is restricted, void volume and FVE distribution dominates, which is why $H_2$ diffusion is higher in TRP, as it has larger voids at 100 K (Figure 3). Secondly, since PMP chains have greater flexibility, it causes the FVEs to dilate, allowing the penetrant to hop more easily between the dilated voids. This is in line with the diffusion theories, which state that penetrant diffusion consists of series of jumps between FVEs in the polymer matrix.[13,73,74] This coupling between penetrant motion and polymer local dynamics has also been observed in other systems, such as phenol diffusion in bisphenol-A-polycarbonate and helium diffusion in polypropylene.[75,76] The mobility of hydrogen is strongly correlated with chain dynamics, and not just the distribution of FVEs.

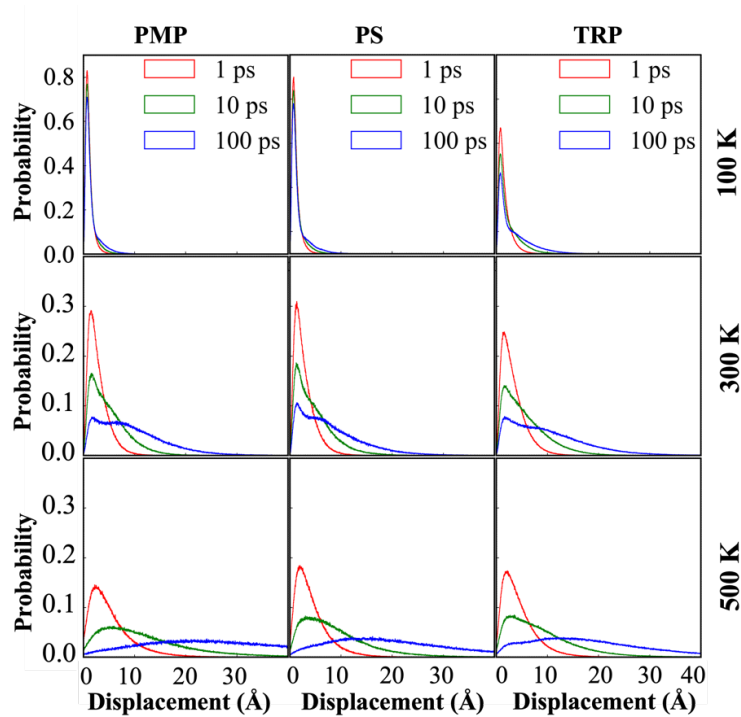

**Figure 10:** Van Hove correlation function showing the distribution of hydrogen gas displacement in PMP, PS and TRP at 100 K, 300 K and 500 K, as labeled. Data shown for three time intervals (dt) of 1 ps, 10 ps, and 100 ps.

**Conclusion**

In summary, we implemented MD simulations to highlight the interplay between temperature, local chain dynamics, free volume element distribution, and penetrant transport in amorphous polymers. By using geometric and energetic methods to calculate the size and FVE distribution of different systems at different temperatures, we were able to probe the stability of FVE distribution against backbone flexibility and polymer segmental motions. We also calculated the dynamics of polymer segments as a function of distance from the voids in the matrix. As polymer segments near the surface have more degrees of orientational freedom, they exhibit faster bond decorrelation compared to bulk segments. With the increase in temperature within a given system, there is an increase in local segmental dynamics, leading to a redistribution of FVEs. The expansion and coalescence of FVEs are evidenced by a shift of FVE distribution towards higher diameter values, coupled with a reduction in the number of smaller FVEs. Upon the introduction of hydrogen penetrant, polymer mobility, in addition to FVE distribution, plays a critical role in the diffusion trends observed across different systems and temperatures. The diffusion of hydrogen in PMP exhibits a significant increase with increasing temperatures, even though its microporosity is lower than PS and TRP. This phenomenon is attributed to assisted transport of polymer chains, where the motion of polymers contributes to the enhancement of hydrogen diffusion.

This work suggests that the relaxation of polymer segments, particularly surface segments, can serve as a useful method for assessing the stability of FVE distribution in different polymer membranes. The mobility of surface segments is directly correlated with the FVE redistribution in the polymer. The impact of backbone flexibility on the stability of FVE distribution is significant, as it affects emergent phenomena such as plasticization and physical aging. Using segmental dynamics to establish a direct correlation between the polymer flexibility and FVE distribution can guide the design of superior membrane materials to address these challenges. Furthermore, while designing polymers for separation applications, it is imperative to decouple the contributions of polymer mobility and void distribution on penetrant transport. Future studies will be carried out to understand how tunable parameters such as polymer chemistry, polydispersity, and crosslinking influence physical aging.

**Acknowledgements**

The authors acknowledge funding provided by the U.S. Department of Energy (DOE), Office of Basic Energy Sciences, Division of Chemical Sciences, Geosciences and Biosciences under award DE-FG02-17ER16362. Part of this research was financially supported by Prof. Sampath's startup funds provided by the Department of Chemical Engineering and Herbert Wertheim College of Engineering at the University of Florida. https://www.che.ufl.edu The authors acknowledge University of Florida Research Computing for providing computational resources and support that have contributed to the research results reported in this publication. http://researchcomputing.ufl.edu. The authors also thank Dylan M. Anstine for many helpful discussions on the system setup.

**Supporting Information**

Details related to polymer parameterization and structure generation, glass transition temperature calculation, and hydrogen-polymer radial distribution function can be found in the Supporting Information.

**Data availability**

Analysis codes, input scripts, and data file containing force field parameters and structures can be found here - https://github.com/UFSRG/published-work/tree/main/2022-Otmi-Macromolecules

# Bibliography

(1)    Sholl, D. S.; Lively, R. P. Seven Chemical Separations to Change the World. *Nature* **2016**, *532* (7600), 435–437.

(2)    Robeson, L. M. The Upper Bound Revisited. *J. Memb. Sci.* **2008**, *320* (1–2), 390–400.

(3)    Galizia, M.; Chi, W. S.; Smith, Z. P.; Merkel, T. C.; Baker, R. W.; Freeman, B. D. *50th Anniversary Perspective* : Polymers and Mixed Matrix Membranes for Gas and Vapor Separation: A Review and Prospective Opportunities. *Macromolecules* **2017**, *50* (20), 7809–7843.

(4)    Kim, S.; Lee, Y. M. Rigid and Microporous Polymers for Gas Separation Membranes. *Prog. Polym. Sci.* **2015**, *43*, 1–32.

(5)    Kadirkhan, F.; Goh, P. S.; Ismail, A. F.; Wan Mustapa, W. N. F.; Halim, M. H. M.; Soh, W. K.; Yeo, S. Y. Recent Advances of Polymeric Membranes in Tackling Plasticization and Aging for Practical Industrial CO2/CH4 Applications-A Review. *Membranes (Basel)* **2022**, *12* (1).

(6)    Swaidan, R.; Ghanem, B.; Litwiller, E.; Pinnau, I. Physical Aging, Plasticization and Their Effects on Gas Permeation in "Rigid" Polymers of Intrinsic Microporosity. *Macromolecules* **2015**, *48* (18), 6553–6561.

(7)    Ismail, A. F.; Lorna, W. Penetrant-Induced Plasticization Phenomenon in Glassy Polymers for Gas Separation Membrane. *Separation and Purification Technology* **2002**, *27* (3), 173–194.

(8)    Wang, X.-Y.; Willmore, F. T.; Raharjo, R. D.; Wang, X.; Freeman, B. D.; Hill, A. J.; Sanchez, I. C. Molecular Simulations of Physical Aging in Polymer Membrane Materials. *J. Phys. Chem. B* **2006**, *110* (33), 16685–16693.

(9)    Guiver, M. D. Polymer Rigidity Improves Microporous Membranes. *Science* **2013**.

(10)   Hoffman, D. J.; Fica-Contreras, S. M.; Fayer, M. D. Amorphous Polymer Dynamics and Free Volume Element Size Distributions from Ultrafast IR Spectroscopy. *Proc Natl Acad Sci USA* **2020**, *117* (25), 13949–13958.

(11)   Fica-Contreras, S. M.; Hoffman, D. J.; Pan, J.; Liang, C.; Fayer, M. D. Free Volume Element Sizes and Dynamics in Polystyrene and Poly(Methyl Methacrylate) Measured with Ultrafast Infrared Spectroscopy. *J. Am. Chem. Soc.* **2021**, *143* (9), 3583–3594.

(12)   Iyer/Liu/Zhang, G. M., Lu ,. hen. Hydrocarbon Separations by Glassy Polymer Membranes. *JOURNAL OF POLYMER SCIENCE* **2020**.

(13)   Cohen, M. H.; Turnbull, D. Molecular Transport in Liquids and Glasses. *J. Chem. Phys.* **1959**, *31* (5), 1164–1169.

(14)   Robeson, L. M.; Liu, Q.; Freeman, B. D.; Paul, D. R. Comparison of Transport Properties of Rubbery and Glassy Polymers and the Relevance to the Upper Bound Relationship. *J. Memb. Sci.* **2015**, *476*, 421–431.

(15)   Matteucci, S.; Yampolskii, Y.; Freeman, B. D.; Pinnau, I. Transport of Gases and Vapors in Glassy and Rubbery Polymers. In *Materials science of membranes for gas and vapor separation*;

Yampolskii, Y., Pinnau, I., Freeman, B., Eds.; John Wiley & Sons, Ltd: Chichester, UK, 2006; pp 1–47.

(16)   Robeson, L. M.; Smith, Z. P.; Freeman, B. D.; Paul, D. R. Contributions of Diffusion and Solubility Selectivity to the Upper Bound Analysis for Glassy Gas Separation Membranes. *J. Memb. Sci.* **2014**, *453*, 71–83.

(17)   Ramesh, N.; Davis, P. K.; Zielinski, J. M.; Danner, R. P.; Duda, J. L. Application of Free-Volume Theory to Self Diffusion of Solvents in Polymers below the Glass Transition Temperature: A Review. *J. Polym. Sci. B Polym. Phys.* **2011**, *49* (23), 1629–1644.

(18)   Thran, A.; Kroll, G.; Faupel, F. Correlation between Fractional Free Volume and Diffusivity of Gas Molecules in Glassy Polymers. *J. Polym. Sci. B Polym. Phys.* **1999**, *37* (23), 3344–3358.

(19)   Park, J. Y.; Paul, D. R. Correlation and Prediction of Gas Permeability in Glassy Polymer Membrane Materials via a Modified Free Volume Based Group Contribution Method. *J. Memb. Sci.* **1997**, *125* (1), 23–39.

(20)   Horn, N. R. A Critical Review of Free Volume and Occupied Volume Calculation Methods. *J. Memb. Sci.* **2016**, *518*, 289–294.

(21)   Heuchel, M.; Fritsch, D.; Budd, P. M.; McKeown, N. B.; Hofmann, D. Atomistic Packing Model and Free Volume Distribution of a Polymer with Intrinsic Microporosity (PIM-1). *J. Memb. Sci.* **2008**, *318* (1–2), 84–99.

(22)   Konnertz, N.; Ding, Y.; Harrison, W. J.; Budd, P. M.; Schönhals, A.; Böhning, M. Molecular Mobility of the High Performance Membrane Polymer PIM-1 as Investigated by Dielectric Spectroscopy. *ACS Macro Lett.* **2016**, *5* (4), 528–532.

(23)   Paul, W.; Bedrov, D.; Smith, G. D. Glass Transition in 1,4-Polybutadiene: Mode-Coupling Theory Analysis of Molecular Dynamics Simulations Using a Chemically Realistic Model. *Phys. Rev. E Stat. Nonlin. Soft Matter Phys.* **2006**, *74* (2 Pt 1), 021501.

(24)   Lyulin, A. V.; Balabaev, N. K.; Michels, M. A. J. Correlated Segmental Dynamics in Amorphous Atactic Polystyrene:  A Molecular Dynamics Simulation Study. *Macromolecules* **2002**, *35* (25), 9595–9604.

(25)   Petropoulos, J. H. Mechanisms and Theories for Sorption and Diffusion of Gases in Polymers. In *Polymeric gas separation membranes*; Paul, D. R., Yampol'skii, Y. P., Eds.; CRC Press, 2018; pp 17–81.

(26)   Low, Z.-X.; Budd, P. M.; McKeown, N. B.; Patterson, D. A. Gas Permeation Properties, Physical Aging, and Its Mitigation in High Free Volume Glassy Polymers. *Chem. Rev.* **2018**, *118* (12), 5871–5911.

(27)   Yu, H. J.; Chan, C.-H.; Nam, S. Y.; Kim, S.-J.; Yoo, J. S.; Lee, J. S. Thermally Cross-Linked Ultra-Robust Membranes for Plasticization Resistance and Permeation Enhancement – A Combined Theoretical and Experimental Study. *J. Memb. Sci.* **2022**, *646*, 120250.

(28)   Zhang, L.; Xiao, Y.; Chung, T.-S.; Jiang, J. Mechanistic Understanding of $CO_2$-Induced Plasticization of a Polyimide Membrane: A Combination of Experiment and Simulation Study. *Polymer* **2010**, *51* (19), 4439–4447.

(29)    Gidley, D. W.; Peng, H.-G.; Vallery, R. S. POSITRON ANNIHILATION AS A METHOD TO CHARACTERIZE POROUS MATERIALS. *Annu. Rev. Mater. Res.* **2006**, *36* (1), 49–79.

(30)    Golemme, G.; Nagy, J. B.; Fonseca, A.; Algieri, C.; Yampolskii, Yu. 129Xe-NMR Study of Free Volume in Amorphous Perfluorinated Polymers: Comparsion with Other Methods. *Polymer* **2003**, *44* (17), 5039–5045.

(31)    Victor, J. G.; Torkelson, J. M. On Measuring the Distribution of Local Free Volume in Glassy Polymers by Photochromic and Fluorescence Techniques. *Macromolecules* **1987**, *20* (9), 2241–2250.

(32)    Kupgan, G.; Demidov, A. G.; Colina, C. M. Plasticization Behavior in Polymers of Intrinsic Microporosity (PIM-1): A Simulation Study from Combined Monte Carlo and Molecular Dynamics. *J. Memb. Sci.* **2018**, *565*, 95–103.

(33)    Kupgan, G.; Abbott, L. J.; Hart, K. E.; Colina, C. M. Modeling Amorphous Microporous Polymers for CO2 Capture and Separations. *Chem. Rev.* **2018**, *118* (11), 5488–5538.

(34)    Anstine, D. M.; Tang, D.; Sholl, D. S.; Colina, C. M. Adsorption Space for Microporous Polymers with Diverse Adsorbate Species. *npj Comput. Mater.* **2021**, *7* (1), 53.

(35)    Hart, K. E.; Colina, C. M. Estimating Gas Permeability and Permselectivity of Microporous Polymers. *J. Memb. Sci.* **2014**, *468*, 259–268.

(36)    Meng, D.; Zhang, K.; Kumar, S. K. Size-Dependent Penetrant Diffusion in Polymer Glasses. *Soft Matter* **2018**, *14* (21), 4226–4230.

(37)    Zhang, K.; Meng, D.; Müller-Plathe, F.; Kumar, S. K. Coarse-Grained Molecular Dynamics Simulation of Activated Penetrant Transport in Glassy Polymers. *Soft Matter* **2018**, *14* (3), 440–447.

(38)    Barnett, J. W.; Kumar, S. K. Modeling Gas Transport in Polymer-Grafted Nanoparticle Membranes. *Soft Matter* **2019**, *15* (3), 424–432.

(39)    Shen, M.; Keten, S.; Lueptow, R. M. Dynamics of Water and Solute Transport in Polymeric Reverse Osmosis Membranes via Molecular Dynamics Simulations. *J. Memb. Sci.* **2016**, *506*, 95–108.

(40)    Xu, Q.; Jiang, J. Molecular Simulations of Liquid Separations in Polymer Membranes. *Curr. Opin. Chem. Eng.* **2020**, *28*, 66–74.

(41)    Frentrup, H.; Hart, K. E.; Colina, C. M.; Müller, E. A. In Silico Determination of Gas Permeabilities by Non-Equilibrium Molecular Dynamics: CO2 and He through PIM-1. *Membranes (Basel)* **2015**, *5* (1), 99–119.

(42)    Brownell, M.; Frischknecht, A. L.; Wilson, M. A. Subdiffusive High-Pressure Hydrogen Gas Dynamics in Elastomers. *Macromolecules* **2022**, *55* (10), 3788–3800.

(43)    Wilson, M. A.; Frischknecht, A. L. High-Pressure Hydrogen Decompression in Sulfur Crosslinked Elastomers. *Int. J. Hydrogen Energy* **2022**, *47* (33), 15094–15106.

(44)   Saha, S.; Bhowmick, A. K. An Insight into Molecular Structure and Properties of Flexible Amorphous Polymers: A Molecular Dynamics Simulation Approach. *J. Appl. Polym. Sci.* **2019**, *136* (18), 47457.

(45)   Smith, G. D.; Smith, W. Structure and Dynamics of Amorphous Polymers: Computer Simulations Compared to Experiment and Theory . *Rep Prog Phys* **2004**.

(46)   Gartner, T. E.; Jayaraman, A. Modeling and Simulations of Polymers: A Roadmap. *Macromolecules* **2019**, *52* (3), 755–786.

(47)   Dodda, L. S.; Cabeza de Vaca, I.; Tirado-Rives, J.; Jorgensen, W. L. LigParGen Web Server: An Automatic OPLS-AA Parameter Generator for Organic Ligands. *Nucleic Acids Res.* **2017**, *45* (W1), W331–W336.

(48)   Abbott, L. J.; Hart, K. E.; Colina, C. M. Polymatic: A Generalized Simulated Polymerization Algorithm for Amorphous Polymers. *Theor. Chem. Acc.* **2013**, *132* (3), 1334.

(49)   Larsen, G. S.; Lin, P.; Hart, K. E.; Colina, C. M. Molecular Simulations of PIM-1-like Polymers of Intrinsic Microporosity. *Macromolecules* **2011**, *44* (17), 6944–6951.

(50)   Karayiannis, N. Ch.; Mavrantzas, V. G.; Theodorou, D. N. Detailed Atomistic Simulation of the Segmental Dynamics and Barrier Properties of Amorphous Poly(Ethylene Terephthalate) and Poly(Ethylene Isophthalate). *Macromolecules* **2004**, *37* (8), 2978–2995.

(51)   Polymer Properties Database.

(52)   Liu, Q.; Borjigin, H.; Paul, D. R.; Riffle, J. S.; McGrath, J. E.; Freeman, B. D. Gas Permeation Properties of Thermally Rearranged (TR) Isomers and Their Aromatic Polyimide Precursors. *J. Memb. Sci.* **2016**, *518*, 88–99.

(53)   Bejagam, K. K.; Iverson, C. N.; Marrone, B. L.; Pilania, G. Molecular Dynamics Simulations for Glass Transition Temperature Predictions of Polyhydroxyalkanoate Biopolymers. *Phys. Chem. Chem. Phys.* **2020**, *22* (32), 17880–17889.

(54)   Mohammadi, M.; fazli, H.; karevan, M.; Davoodi, J. The Glass Transition Temperature of PMMA: A Molecular Dynamics Study and Comparison of Various Determination Methods. *Eur. Polym. J.* **2017**, *91*, 121–133.

(55)   Pilgrim, C. Piecewise-Regression (Aka Segmented Regression) in Python. *JOSS* **2021**, *6* (68), 3859.

(56)   Thompson, A. P.; Aktulga, H. M.; Berger, R.; Bolintineanu, D. S.; Brown, W. M.; Crozier, P. S.; in 't Veld, P. J.; Kohlmeyer, A.; Moore, S. G.; Nguyen, T. D.; et al. LAMMPS - a Flexible Simulation Tool for Particle-Based Materials Modeling at the Atomic, Meso, and Continuum Scales. *Comput. Phys. Commun.* **2022**, *271*, 108171.

(57)   White, R. P.; Lipson, J. E. G. Polymer Free Volume and Its Connection to the Glass Transition. *Macromolecules* **2016**, *49* (11), 3987–4007.

(58)   Sarkisov, L.; Harrison, A. Computational Structure Characterisation Tools in Application to Ordered and Disordered Porous Materials. *Mol. Simul.* **2011**, *37* (15), 1248–1257.

(59) T.Willmore, F.; Wang, X.; Sanchez, I. Free Volume Properties of Model Fluids and Polymers: Shape and Connectivity. *Wiley Interdiscip Rev Comput Mol Sci* **2004**, No. December 2005, 1385–1393.

(60) Willems, T. F.; Rycroft, C. H.; Kazi, M.; Meza, J. C.; Haranczyk, M. Algorithms and Tools for High-Throughput Geometry-Based Analysis of Crystalline Porous Materials. *Micropor. Mesopor. Mat.* **2012**, *149* (1), 134–141.

(61) Sarkisov, L.; Bueno-Perez, R.; Sutharson, M.; Fairen-Jimenez, D. Materials Informatics with PoreBlazer v4.0 and the CSD MOF Database. *Chem. Mater.* **2020**, *32* (23), 9849–9867.

(62) Stukowski, A. Visualization and Analysis of Atomistic Simulation Data with OVITO–the Open Visualization Tool. *Modelling Simul. Mater. Sci. Eng.* **2010**, *18* (1), 015012.

(63) Stukowski, A. Computational Analysis Methods in Atomistic Modeling of Crystals. *JOM* **2014**, *66* (3), 399–407.

(64) Yampolskii, Y. P. Methods for Investigation of the Free Volume in Polymers. *Russ. Chem. Rev.* **2007**, *76* (1), 59–78.

(65) Jiang, Y.; Willmore, F. T.; Sanders, D.; Smith, Z. P.; Ribeiro, C. P.; Doherty, C. M.; Thornton, A.; Hill, A. J.; Freeman, B. D.; Sanchez, I. C. Cavity Size, Sorption and Transport Characteristics of Thermally Rearranged (TR) Polymers. *Polymer* **2011**, *52* (10), 2244–2254.

(66) Willmore, F. T. A Toolkit for the Analysis and Visualization of Free Volume in Materials. In *Proceedings of the 1st Conference of the Extreme Science and Engineering Discovery Environment on Bridging from the eXtreme to the campus and beyond - XSEDE '12*; ACM Press: New York, New York, USA, 2012; p 1.

(67) Willmore, F. T.; Wang, X.; Sanchez, I. C. Free Volume Properties of Model Fluids and Polymers: Shape and Connectivity. *J. Polym. Sci. B Polym. Phys.* **2006**, *44* (9), 1385–1393.

(68) in 't Veld, P. J.; Stone, M. T.; Truskett, T. M.; Sanchez, I. C. Liquid Structure via Cavity Size Distributions. *J. Phys. Chem. B* **2000**, *104* (50), 12028–12034.

(69) Lock, S. S. M.; Lau, K. K.; Mei, I. L. S.; Shariff, A. M.; Yeong, Y. F. Cavity Energetic Sizing Algorithm Applied in Polymeric Membranes for Gas Separation. *Procedia Engineering* **2016**, *148*, 855–861.

(70) Persistence of Vision Pty. Ltd http://www.povray.org/download/ (accessed Mar 27, 2023).

(71) Abdel-Hady, E. E.; Mohamed, H. F. M.; El-Sharkawy, M. R. M. Temperature Effect on Free Volume of Polymethylpentene Studied by Positron Annihilation Technique. *phys. stat. sol. (c)* **2009**, *6* (11), 2420–2422.

(72) Lyulin, A. V.; Michels, M. A. J. Molecular Dynamics Simulation of Bulk Atactic Polystyrene in the Vicinity of $t_g$. *Macromolecules* **2002**, *35* (4), 1463–1472.

(73) Pace, R. J.; Datyner, A. Statistical Mechanical Model for Diffusion of Simple Penetrants in Polymers. I. Theory. *J. Polym. Sci. Polym. Phys. Ed.* **1979**, *17* (3), 437–451.

(74)   DiBenedetto, A. T.; Paul, D. R. An Interpretation of Gaseous Diffusion through Polymers Using Fluctuation Theory. *J. Polym. Sci. A Gen. Pap.* **1964**, *2* (2), 1001–1015.

(75)   Hahn, O.; Mooney, D. A.; Müller-Plathe, F.; Kremer, K. A New Mechanism for Penetrant Diffusion in Amorphous Polymers: Molecular Dynamics Simulations of Phenol Diffusion in Bisphenol-A-Polycarbonate. *J. Chem. Phys.* **1999**, *111* (13), 6061–6068.

(76)   Boshoff, J. H. D.; Lobo, R. F.; Wagner, N. J. Influence of Polymer Motion, Topology and Simulation Size on Penetrant Diffusion in Amorphous, Glassy Polymers:  Diffusion of Helium in Polypropylene. *Macromolecules* **2001**, *34* (17), 6107–6116.